# Influence of the Dynamic Social Network Timeframe Type and Size on the Group Evolution Discovery

Stanisław Saganowski, Piotr Bródka, Przemysław Kazienko
Institute of Informatics, Wrocław University of Technology, Wyb.Wyspiańskiego 27, 50-370 Wrocław, Poland
stanislaw.saganowski@pwr.wroc.pl, piotr.brodka@pwr.wroc.pl, kazienko@pwr.wroc.pl

*Abstract*—New technologies allow to store vast amount of data about users interaction. From those data the social network can be created. Additionally, because usually also time and dates of this activities are stored, the dynamic of such network can be analysed by splitting it into many timeframes representing the state of the network during specific period of time. One of the most interesting issue is group evolution over time. To track group evolution the GED method can be used. However, choice of the timeframe type and length might have great influence on the method results. Therefore, in this paper, the influence of timeframe type as well as timeframe length on the GED method results is extensively analysed.

## I. INTRODUCTION AND RELATED WORK

In modern telecommunication systems, the relationships between users are discovered based on system logs, containing information on the elementary events - usually relating to services considered system (message e-mail, phone call, etc..). Events are discrete, however the relationships used in further analysis of the social network are continuous or relate to the selected timeframe. The consequence of this fact is the variability of network structure.

This factor is a major problem in analysis of dynamic in social networks. Numerous studies indicate that the values of basic parameters (the node degree, node centrality, the composition of groups, etc..) calculated for successive periods (timeframes) show high variability and lack of correlation [1].

In [1] authors proved that for successive, separated timeframes, the correlation between network structure or node degree is very low. Therefore it is very hard to analyse such network.

Similar results was presented in [2] and [3]. When the timeframe is too narrow, there is a lack of correlation and a big noise in the structural parameters of the network, while a large time window (in the extreme case including the data from the entire available period) leads to loss of information on the temporal relationship between the analysed network connectivity. Additionally for different structural measures the appropriate size and type of the time window may be different, which clearly leads to the conclusion that the dynamic analysis of social network is a very hard task [4], [5].

The GED method [6] to extract group evolution history utilize the social network in form of successive timeframes, so the results of the method may depend on the selected social network division into timeframes. Thus in this paper, the influence of timeframe type as well as timeframe size on the group evolution extraction is analysed.

## II. GROUP EVOLUTION DISCOVERY

Before the method can be presented, it is necessary to describe a few concepts related to social networks

### A. Temporal Social Network and Groups

Temporal social network *TSN* is a list of following timeframes (time windows) *T*. Each timeframe is in fact one social network $SN(V,E)$ where $V$ – is a set of vertices and $E$ is a set of directed edges $<x,y>:x,y \in V, x \neq y$

$$\begin{aligned}
TSN &= <T_1, T_2, ...., T_m>, \quad m \in N \\
T_i &= SN_i(V_i, E_i), \quad i = 1, 2, ..., m \\
E_i &= <x, y>: x, y \in V_i, x \neq y \quad i = 1, 2, ..., m
\end{aligned} \quad (1)$$

There is no universally acceptable definition of the group (social community) in social networks [7]. There are several of them, which are used depending on the authors' needs. In addition, some of them cannot be even called definitions but only criteria for the group existence.

A group, often also called a social community, in the biological terminology is a number of cooperating organisms, sharing a common environment. In sociology, in turn, it is usually defined as a group of people living and cooperating in a single location. However, due to the rapid growth of Internet, the concept of community has lost its geographical limitations. Overall, a general idea of the social community is a set of people in a social network, whose members more frequently collaborate with each other rather than with members of this social network who do not belong to the group. This concept of the social community can be easily transposed to the graph theory, in which the social network is represented by a graph. Group is a set of vertices with high density of edges between them, and low edge density between those vertices and nodes which do not belong to this set. However, the problem arises in the quantitative definition of community. Most definitions are build based on the idea presented above. Nevertheless, as mentioned earlier, there are many alternative approaches and none of them has been commonly accepted. Additionally,

The work was partly supported by the Fellowship co-Financed by European Union within European Social Fund and Polish National Science Centre - the research project, 2010-2013.

groups can also be algorithmically determined, as the output of the specific clustering algorithm, i.e. without a precise a priori definition [8]. In this paper, we will use such definition, i.e. a group $G$ extracted from the social network $SN(V,E)$ is a subset of vertices from $V$ ($G \subseteq V$), extracted using any community extraction method (clustering algorithm).

### B. Group Evolution

Evolution of particular social community can be represented as a sequence of events (changes) following each other in the successive time windows (timeframes) within the temporal social network. Possible events in social group evolution are:

1. *Continuing* (stagnation) – the group continue its existence when two groups in the consecutive time windows are identical or when two groups differ only by few nodes but their size remains the same.

2. *Shrinking* – the group shrinks when some nodes has left the group, making its size smaller than in the previous time window. Group can shrink slightly i.e. by a few nodes or greatly losing most of its members.

3. *Growing* (opposite to shrinking) – the group grows when some new nodes have joined the group, making its size bigger than in the previous time window. A group can grow slightly as well as significantly, doubling or even tripling its size.

4. *Splitting* – the group splits into two or more groups in the next time window when few groups from timeframe $T_{i+1}$ consist of members of one group from timeframe $T_i$. We can distinguish two types of splitting: (1) equal, which means the contribution of the groups in split group is almost the same and (2) unequal when one of the groups has much greater contribution in the split group, which for this one group the event might be similar to shrinking.

5. *Merging* (reverse to splitting) – the group has been created by merging several other groups when one group from timeframe $T_{i+1}$ consist of two or more groups from the previous timeframe $T_i$. Merge, just like the split, might be (1) equal, which means the contribution of the groups in merged group is almost the same, or (2) unequal, when one of the groups has much greater contribution into the merged group.

6. *Dissolving* happens when a group ends its life and does not occur in the next time window, i.e., its members have vanished or stop communicating with each other and scattered among the rest of the groups.

7. *Forming* (opposed to dissolving) of new group occurs when group which has not existed in the previous time window $T_i$ appears in next time window $T_{i+1}$.

### C. GED – a Method for Group Evolution Discovery in the Social Network

To track social community evolution in social network the new method called *GED* (Group Evolution Discovery) was developed. Key element of this method is a new measure called inclusion. This measure allows to evaluate the inclusion of one group in another. Therefore, inclusion of group $G_1$ in group $G_2$ is calculated as follows:

$$I(G_1, G_2) = \overbrace{\frac{|G_1 \cap G_2|}{|G_1|}}^{group\ quantity} \cdot \underbrace{\frac{\sum_{x \in (G_1 \cap G_2)} NI_{G_1}(x)}{\sum_{x \in (G_1)} NI_{G_1}(x)}}_{group\ quality} \quad (2)$$

where $NI_{G_1}(x)$ is the value reflecting importance of the node $x$ in group $G_1$.

As a node importance measure, any metric which indicate member position within the community can be used, e.g. centrality degree, betweenness degree, page rank, social position etc. The second factor in Equation 2 would have to be adapted accordingly to selected measure.

As mentioned earlier the *GED* method, used to track group evolution, takes into account both the quantity and quality of the group members. The quantity is reflected by the first part of the inclusion measure, i.e. what portion of $G_1$ members is shared by both groups $G_1$ and $G_2$, whereas the quality is expressed by the second part of the inclusion measure, namely what contribution of important members of $G_1$ is shared by both groups $G_1$ and $G_2$. It provides a balance between the groups, which contain many of the less important members and groups with only few but key members.

It is assumed that only one event may occur between two groups ($G_1$, $G_2$) in the consecutive timeframes, however one group in timeframe $T_i$ may have several events with different groups in $T_{i+1}$.

The indicators $\alpha$ and $\beta$ are the *GED* method parameters which can be used to adjust the method to particular social network and community detection method. After the experiments in [6] authors suggest that the values of $\alpha$ and $\beta$ should be from range [50%; 100%]

Based on the list of extracted events, which have been extracted by GED method for selected group between each two successive timeframes, the group evolution is created. In the example presented in Figure 1 the network consists from eight time windows. The group forms in $T_2$, then by gaining new nodes grows in $T_3$, next splits into two groups in $T_4$, then by losing one node the bigger group is shrinking in $T_5$, both groups continue over $T_6$, next both groups merges with the third group in $T_7$, and finally the group dissolves in $T_8$.

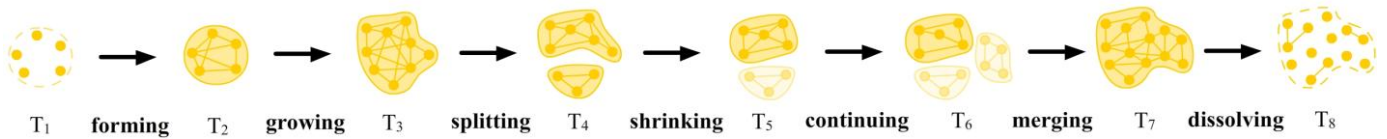

Figure 1. Changes over time for the single group.

---

*GED* – Group Evolution Discovery Method

**Input:** *TSN* in which at each timeframe $T_i$ groups are extracted by any community detection algorithm. Calculated any user importance measure.

1. For each pair of groups <$G_1$, $G_2$> in consecutive timeframes $T_i$ and $T_{i+1}$ inclusion of $G_1$ in $G_2$ and $G_2$ in $G_1$ is counted according to equation (2).
2. Based on inclusion and size of two groups one type of event may be assigned:

   a. *Continuing*: $I(G_1,G_2) \geq \alpha$ and $I(G_2,G_1) \geq \beta$ and $|G_1| = |G_2|$

   b. *Shrinking*: $I(G_1,G_2) \geq \alpha$ and $I(G_2,G_1) \geq \beta$ and $|G_1| > |G_2|$ OR $I(G_1,G_2) < \alpha$ and $I(G_2,G_1) \geq \beta$ and $|G_1| \geq |G_2|$ OR $I(G_1,G_2) \geq \alpha$ and $I(G_2,G_1) < \beta$ and $|G_1| \geq |G_2|$ and there is only one match between $G_1$ and groups in the next time window $T_{i+1}$

   c. *Growing*: $I(G_1,G_2) \geq \alpha$ and $I(G_2,G_1) \geq \beta$ and $|G_1|<|G_2|$ OR $I(G_1,G_2) \geq \alpha$ and $I(G_2,G_1) < \beta$ and $|G_1| \leq |G_2|$ OR $I(G_1,G_2) < \alpha$ and $I(G_2,G_1) \geq \beta$ and $|G_1| \leq |G_2|$ and there is only one match between $G_2$ and groups in the next previous window $T_i$

   d. *Splitting*: $I(G_1,G_2) < \alpha$ and $I(G_2,G_1) \geq \beta$ and $|G_1| \geq |G_2|$ OR $I(G_1,G_2) \geq \alpha$ and $I(G_2,G_1) < \beta$ and $|G_1| \geq |G_2|$ and there is more than one match between $G_1$ and groups in the next time window $T_{i+1}$

   e. *Merging*: $I(G_1,G_2) \geq \alpha$ and $I(G_2,G_1) < \beta$ and $|G_1| \leq |G_2|$ OR $I(G_1,G_2) < \alpha$ and $I(G_2,G_1) \geq \beta$ and $|G_1| \leq |G_2|$ and there is more than one match between $G_2$ and groups in the previous time window $T_i$

   f. *Dissolving*: for $G_1$ in $T_i$ and each group $G_2$ in $T_{i+1}$ $I(G_1,G_2) < 10\%$ and $I(G_2,G_1) < 10\%$

   g. *Forming*: for $G_2$ in $T_{i+1}$ and each group $G_1$ in $T_i$ $I(G_1,G_2) < 10\%$ and $I(G_2,G_1) < 10\%$

---

## III. EXPERIMENTS

### A. Data Set

Data utilized in the experiments were obtained from the portal extradom.pl. It gathers people, who are engaged in building their own houses in Poland. It helps them to exchange best practices, experiences, evaluate various constructing projects and technologies or simply to find the answers to their questions provided by others. The data covers a period of 17 months and contains 3,690 users and 34,082 relations. The nodes in the social network represent the users of extradom.pl and the edges activities between them, i.e. private messages, forum posts, photos exchange.

In the experiment three types of timeframe were studied: (1) disjoint– the end of one time window is the beginning of next one, (2) overlapping– offset in days of the consecutive time windows is shorter than time window size, so the following time window starts before the previous ends, e.g. the first timeframe begins on the 1st day and ends on the 60th day, second begins on the 31th day and ends on the 90th day and so on (timeframe size is 60 days and offset 30 days), (3) increasing– following time window include all previous time windows, thus its size is bigger with each timeframe.

For disjoint timeframes three sizes were selected: (1) size of 30 days and 30 days' offset (s30o30), (2) size of 60 days and 60 days' offset (s60o60), (3) 90 days' timeframe and 90 days' offset (s90o90). For overlapping timeframes five sizes were chosen: (1) size of 30 days and 15 days' offset (s30o15), (2) size of 60 days and 30 days' offset (s60o30), (3) size of 90 days and 30 days' offset (s90o30), (4) size of 180 days and 30 days' offset (s180o30), (5) 90 days' timeframe and 60 days' offset (s90o60).

For increasing timeframes offset was 30 days (s_o30), so the first timeframe begins on the 1st day and ends on the 30th day, second begins on the 1st day and ends on the 60th day and so on.

For group extraction the CPM clustering method implemented in CFinder (www.http://cfinder.org/) was utilized. The groups were discovered for clique size of 5 nodes and for the directed and unweighted social network. As a node importance measure *social position* was used [9], [10]. The results are presented in Table 1 and Figure 2.

TABLE 1. THE RESULTS OF GED COMPUTATION FOR DIFFERENT TIMEFRAME TYPES

| Timeframe | | | No. Of | | Avg. group size | No. of events for alpha and beta equals 70% | | | | | | | | No. of events for thresholds from range [50%; 100%] |
|---|---|---|---|---|---|---|---|---|---|---|---|---|---|---|
| type | size | offset | timeframes | groups | | Form | Dissolve | Shrink | Growth | Continuation | Split | Merge | Total | |
| disjoint | 30 | 30 | 16 | 287 | 7 | 216 | 190 | 1 | 1 | 0 | 0 | 0 | 408 | 14703 |
| disjoint | 60 | 60 | 8 | 309 | 8 | 207 | 178 | 0 | 3 | 0 | 0 | 0 | 388 | 14007 |
| disjoint | 90 | 90 | 5 | 287 | 9 | 195 | 155 | 0 | 1 | 0 | 0 | 0 | 351 | 12654 |
| overlapping | 90 | 60 | 8 | 470 | 9 | 192 | 165 | 22 | 43 | 11 | 0 | 4 | 437 | 15679 |
| overlapping | 30 | 15 | 33 | 601 | 7 | 222 | 210 | 60 | 72 | 60 | 10 | 4 | 638 | 22784 |
| overlapping | 60 | 30 | 16 | 623 | 8 | 200 | 177 | 59 | 86 | 44 | 0 | 12 | 578 | 20779 |
| overlapping | 90 | 30 | 15 | 908 | 9 | 210 | 190 | 125 | 170 | 129 | 21 | 31 | 876 | 31157 |
| overlapping | 180 | 30 | 14 | 1638 | 10 | 166 | 167 | 296 | 302 | 549 | 71 | 61 | 1612 | 57095 |
| increasing | 30-480 | 30 | 16 | 1965 | 13 | 227 | 0 | 3 | 579 | 1030 | 3 | 122 | 1964 | 70704 |

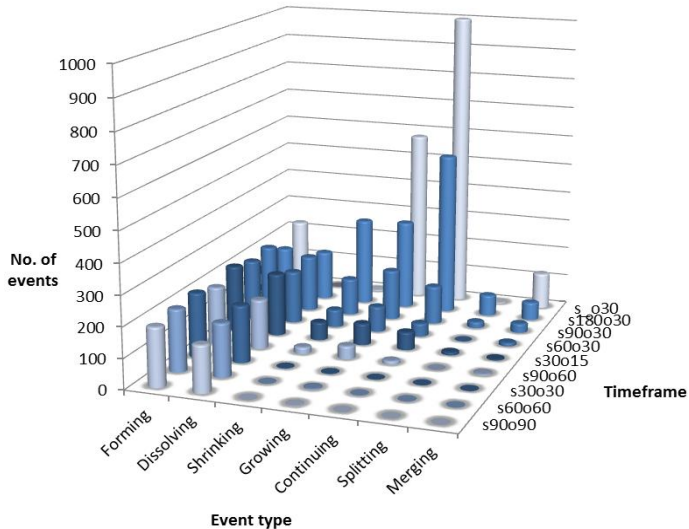

Figure 2. The number of events obtained with the *GED* method run with parameters *α* and *β* equals 70% for different timeframe type and size.

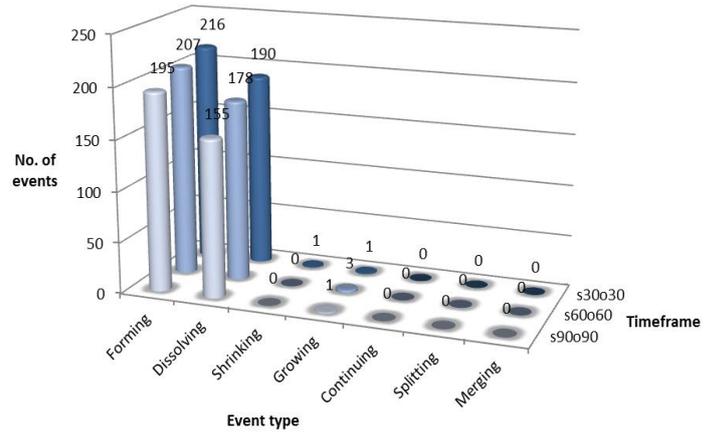

Figure 3. The number of events obtained with the *GED* method run with parameters *α* and *β* equals 70% for disjoint timeframes

### B. Disjoint Timeframes

Comparison of disjoint timeframes is presented in Figure 3, *α* and *β* parameters of the GED method equals 70%. As it can be observed, the GED method provided mostly forming and dissolving events. There are also few growing and shrinking events, but the number is scarce and says nothing about group evolution. The reason for such results is dataset which varies over time. Changes for disjoint timeframes are so rapid that we are not able to catch any *evolution events*. Speaking of the *evolution events* we have in mind process of group changes, while forming and dissolving events are only the beginning and the end of this process.

Increasing size of the timeframe has not improved the number of evolution events, what is more, the number of forming and dissolving events has decreased, as there is less timeframes in the analysed temporal social network. In such situation it is necessary to use overlapping timeframes.

The experiments were repeated for all *α* and *β* parameters from the set {50%, 60%, 70%, 80%, 90%, 100%} and the results were the same.

### C. Overlapping Timeframes

Figure 4 shows the results of the GED method with *α* and *β* parameters equals 70% for different overlapping timeframes. The chart clearly shows that increase of the timeframe size while offset remains the same (extending overlapping), effects in greater number of evolution events, not only forming and dissolving events. This follows from the fact that adjacent timeframes contain some users' interaction from the same period of time and thus changes within network are not so rapid. Therefore, the GED method is able to match groups in consecutive timeframes. It is most visible for the 180 days long timeframe. Because of the large overlapping of timeframes changes between social networks in consecutive time widows are small and there are many continuation events, as well as growing and shrinking events.

Two timeframes were evaluated where size was two times greater than the offset – s30o15 and s60o30. Difference between the number of events found with those timeframes are barely noticeable, however in case of s30o15 timeframe the number of timeframes is twice as in case of s60o30. This implies that the evolution can be investigated more detailed (changes occur within a shorter period of time), but the number of events between two subsequent timeframes is lower (the number of events between all timeframes is almost the same as in case of s60o30 timeframe, but there is more timeframes when using s30o15 timeframe).

The experiments were repeated for all α and β parameters from the set {50%, 60%, 70%, 80%, 90%, 100%} and the conclusion were the same.

### D. Increasing Timeframes

The number of events obtained with increasing timeframes and both parameters equals 70% are presented in Figure 5. Each successive timeframe contains all the relations and nodes from previous timeframes, therefore dissolving events does not appear since the nodes never vanish from the network. From the same reason number of shrinking events is very low and number of continuing events is extremely high, on average 54% of all events are continuing events.

Changes within temporal social network created from increasing timeframes are not very rapid and causes growth of the groups over time, hence the number of discovered events is so great. Such a results might be especially useful for researchers who look for, so called, 'persistent groups' – the groups which lasts over long period of time without significant changes, e.g. turnover of most nodes.

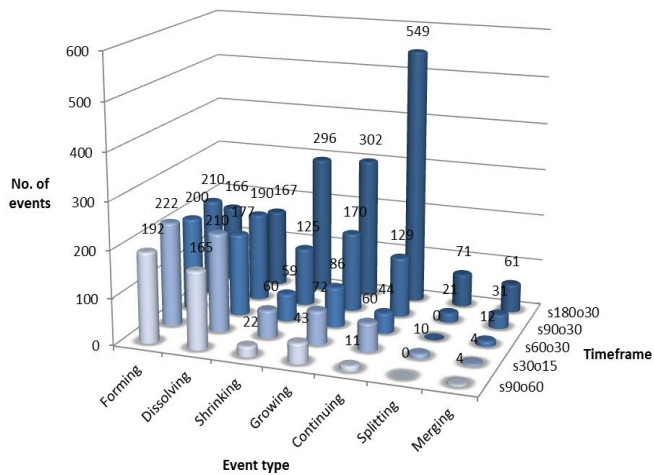

Figure 4. The number of events obtained with the *GED* method run with parameters α and β equals 70% for overlapping timeframes.

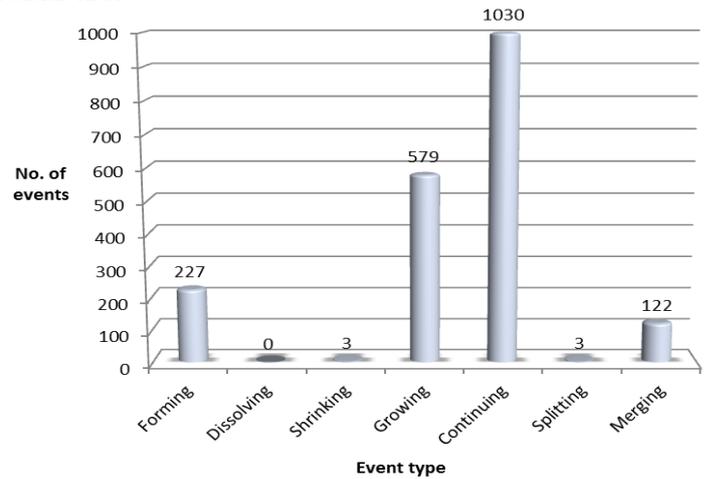

Figure 5. The number of events obtained with the *GED* method run with parameters *α* and *β* equals 70% for increasing timeframe with offset equals 30 days.

The experiments were repeated for all α and β parameters from the set {50%, 60%, 70%, 80%, 90%, 100%} and the number of continuing events was the same. In general the number of all discovered events across different α and β parameters is almost static – it varies from 1951 to 1987 events. This shows that the α and β parameters have very small impact on the number of events found with the increasing timeframe. However, growth of the β parameter reduces the number of growing events for the benefit of merging events.

IV. CONCLUSIONS AND FUTURE WORK

To discover group evolution with the GED method the social network in form of successive timeframes is required. Therefore the results of the method depends on the selected social network division into timeframes. Thus in this paper, the influence of timeframe type as well as timeframe size on the group evolution extraction was analysed. In the experiment three types of timeframe were studied: disjoint, overlapping and increasing. The results are presented in Table 1 and Figure 2.

Changes between successive timeframes in analysed dataset turned out to be too rapid for disjoint timeframes. The results provided mostly forming and dissolving events, which are not sufficient to create full evolution history for the group. Increasing the size of the timeframe has not improve the results.

However, the division into overlapping timeframes were successful and gave great number of diverse events. The number of events was rising while extending the overlapping between adjacent timeframes.

Finally, increasing timeframes provided results attractive for researchers looking for persistent groups, since the GED method with increasing timeframe returned mostly continuing and growing events.

The results shows that for the rapidly changing social networks the overlapping timeframe is the best choice. The type and the size of the timeframe might be additional parameter of the GED method, which can be adjusted. This makes the GED method even more flexible and useful.